# Ectopic expression of *Hoxb1* induces cardiac and craniofacial malformations

Stéphane Zaffran, Gaëlle Odelin, Sonia Stefanovic, Fabienne Lescroart, and Heather C. Etchevers *

Aix Marseille Univ, MMG, INSERM, U1251, Marseille, France

Full address: MMG, INSERM U1250, Faculté de Médecine, Aix-Marseille University, 27 boulevard Jean

Moulin, 13005 Marseille, France

**Running head**: Ectopic Hoxb1 reshapes face and heart

**Keywords**: valve, mandibular agenesis, maxillary aplasia, exencephaly, double outlet right ventricle,

cleft palate

* Corresponding author: heather.etchevers@inserm.fr

## Abstract

Members of the large family of Hox transcription factors are encoded by genes whose tightly regulated expression in development and in space within different embryonic tissues confer positional identity from the neck to the tips of the limbs. Many structures of the face, head and heart develop from cell populations expressing few or no *Hox* genes. *Hoxb1* is the member of its chromosomal cluster expressed in the most rostral domain during vertebrate development, but never by the multipotent neural crest cell population anterior to the cerebellum. We have developed a novel floxed transgenic mouse line, *CAG-Hoxb1,-EGFP* (*CAG-Hoxb1*)*,* which upon recombination by Cre recombinase conditionally induces robust *Hoxb1* and *eGFP* over-expression. When induced within the neural crest lineage, pups die at birth. A variable phenotype develops from E11.5 on, associating frontonasal hypoplasia/aplasia, micrognathia/agnathia, major ocular and forebrain anomalies, and cardiovascular malformations. Neural crest derivatives in the body appear unaffected. Transcription of effectors of developmental signaling pathways (Bmp, Shh, Vegfa) and transcription factors (Pax3, Sox9) is altered in mutants. These outcomes emphasize that repression of *Hoxb1*, along with other paralog group 1 and 2 *Hox* genes, is strictly necessary in anterior cephalic NC for craniofacial, visual, auditory and cardiovascular development.

## Introduction

For over 600 million years of animal evolution, the multiple genes encoding the Hox family of transcription factors have been master conductors of segmental positional identity in the body and limbs (Pearson, Lemons, & McGinnis, 2005). Genomic duplications and rearrangements have led to the presence in most vertebrates of four homologous clusters of thirteen *Hox* genes, of which only 39 have been retained over the years in humans and mice and 38 in chickens, such that each cluster contains between 9 and 11 genes. The clusters of genes exhibit collinearity: according to the physical order of genes within each cluster from 3' to 5', the rostral expression limit of each is globally arranged along the axis from posterior head to tail, spaced according to metameric landmarks of the body and limbs, and the genes are also expressed sequentially in time (Santagati & Rijli, 2003). However, detailed examination in the different germ layers and over time shows that this rule of thumb is not perfectly respected, particularly in the rostral region.

The *Hox* genes expressed most anteriorly in the developing mouse or chick embryo include those of paralog groups 1 and 2. Extant members in these species are *Hoxa1*, *Hoxb1*, and *Hoxd1* (homologous to the Drosophila *labial* gene) and *Hoxa2* and *Hoxb2* (homologous to Drosophila *proboscipedia*). The expression domains of *labial* and *proboscipedia* correspond to rostral structures of the fruitfly head. During the evolution of vertebrates, additional head structures developed both anterior and ventral to the corresponding domains expressing the equivalent paralog groups, comprising not only tissues derived from the three original germ layers but also a major mesenchymal contribution from an evolutionarily new cell population known as the neural crest (NC). The "new head" hypothesis was developed 35 years ago to examine the transition from proto-chordates to the successful vertebrate clade (Gans & Northcutt, 1983). It emphasized that the NC's striking propensity to "function like mesoderm" in the anterior head was the primary element enabling vertebrate characteristics such as active predation, particularly in building the rostral skull vault and facial skeleton, protecting brain and sensory organs, and, for gnathostomes, developing jaws. This entire craniofacial skeleton is derived from NC cells of the first pharyngeal arch (PA) and frontonasal bud (Le Lièvre, 1978, and references therein).

The ectoderm, mesoderm and NC cell mesenchyme surrounding the oral endoderm ventrally in the first PA and joined by the neuroectoderm dorsally are each devoid of *Hox* gene expression. In contrast, the hindbrain neuroectoderm and 2$^{nd}$ and posterior PAs are the sites of the rostral limits of *Hox* genes and are, at least temporarily, visibly metameric. Collinearity in these rostral limits is not strictly observed, either within the derivatives of a given germ layer, or across them in structures of mixed embryological origin, like a PA (Couly, Grapin-Botton, Coltey, Ruhin, & Le Douarin, 1998). In addition, rostral *Hox* expression limits are dynamic, making it difficult to generalize; some exceptions in specific neuronal cell groups can be found depending on the stages examined (Wolf, Yeung, Doucette, & Nazarali, 2001). Nonetheless, the importance of paralog group 1 and 2 *Hox* genes in the identity of specific hindbrain segments (morphologically separated by epithelial constrictions during a brief window as "rhombomeres") is illustrated by several loss-of-function experiments in mice. Null mutation of *Hoxa1* leads to the absence of rhombomere (r)5 and a large part of r4 (Dollé et al., 1993). The *Hoxb1* homozygote mutant mouse fails to specify r4 motor neurons, destined to contribute to the facial motor nucleus in the hindbrain (Goddard, Rossel, Manley, & Capecchi, 1996). Global disruption of the transcription factor gene *Egr2* (also known as *Krox20*), which indirectly represses *Hoxb1* (Garcia-Dominguez, Gilardi-Hebenstreit, & Charnay, 2006) but directly activates *Hoxb2* expression (Sham et al., 1993), leads to reduction or loss of r3 and 5 (Schneider-Maunoury et al., 1993).

*Hoxb1* is strongly expressed in r4 and the spinal cord caudal to r8. Alone, r4 provides the majority of NC mesenchyme to PA2 (contributing to the hyoid, stapes and styloid) and a small amount to PA1 (proximal jaw bones such as the squamosal, palatoquatrate, malleus and incus) (Grapin-Botton, Bonnin, McNaughton, Krumlauf, & Le Douarin, 1995; Köntges & Lumsden, 1996). *Hoxa2* and *Hoxb1* are normally maintained in PA2 NC mesenchyme but downregulated in the small subpopulation of cells from r2-r4 populating the caudal portion of PA1 (Couly et al., 1998; Prince & Lumsden, 1994). *Hoxa2*-homozygous mutant mice acquire mirror-image proximal jaw bones in place of those normally formed from PA2, but no distal or anterior elements such as the prominent Meckel's or palatine cartilages (Rijli et al., 1993). Only NC cells that had once expressed *Hoxb1* are responsive to a change in positional identity in the absence of *Hoxa2*.

How would naïve NC, never having expressed *Hox* genes, respond to their ectopic induction? Previous work has only indirectly addressed this question. Transplantations of quail neural tube segments to more caudal positions in chick hosts, for example from *Hoxb1*-negative r5/6 to a caudal r8 position, come to express *Hoxb1* in the grafted neuroepithelium, while the corresponding NC assume a new, posterior identity by integrating appropriately into the superior ganglion of the vagal nerve. However, r1/r2 neuroepithelium, which never expresses *Hox* genes, is not competent to respond in the same way (Couly et al., 1998; Grapin-Botton et al., 1995). Conversely, where *Hoxb1*-expressing r4 is transplanted to r1-r3 regions, the neuroepithelium maintains its cell-autonomous *Hoxb1* expression, but in half of the manipulated embryos, *Hoxb1* is no longer transcribed in the grafted NC cells (Kuratani & Eichele, 1993), correlating with variable effects on the morphologies of cranial ganglia V to VII and their nerves. Nonetheless, when r4-r6 neuroepithelium is transplanted to the level of the posterior mesencephalon, *Hoxa2* is autonomously maintained in grafted NC cells within the mandibular arch (posterior PA1). Interestingly, this structure becomes hypoplasic by E4, and by E7-E9, all distal jaw and hyoid (in chick, entoglossal or basihyal tongue) skeleton is missing (Couly et al., 1998).

We have generated a conditional transgenic mouse line to examine the consequence of targeted ectopic expression of the *Hoxb1* gene. To revisit and investigate the role of Hoxb1 in patterning of head and neck-derived tissues, we have crossed this new allele with the *Wnt1-Cre* line. *Wnt1-Cre* drives Cre recombinase expression in discrete areas of the presumptive dorsal brain and spinal cord and subsequently, in the majority of NC cells, including those of the *Hox*-negative regions of the head (Chai et al., 2000; Echelard, Vassileva, & McMahon, 1994). The resultant mice demonstrate a range of malformations affecting not only the craniofacial skeleton but also the forebrain, eyes, head and neck glands and the outflow tract of the heart. The milder end of the spectrum is reminiscent of human congenital malformations collectively known as "multiple congenital anomaly" syndromes, suggesting that their common phenotypic associations could stem from induction of temporally or spatially ectopic *HOX* gene expression in the NC lineage during embryonic development.

# Results

## Expression of *CAG-Hoxb1* in neural crest cell lineages causes a range from hypomorphic to severe craniofacial phenotypes

We developed a mouse line, *Tg(CAG-Hoxb1,-EGFP)$^{1Sza}$* (*CAG-Hoxb1*)*,* which upon recombination by Cre recombinase conditionally induces robust *Hoxb1* as well as *eGFP* over-expression (Supplementary Figure 1). To drive expression within most of the cranial as well as all trunk NC lineages, we made use of the recently renamed *H2afv$^{Tg(Wnt1-cre)11Rth}$* (*Wnt1-Cre*) line (Danielian, Muccino, Rowitch, Michael, & McMahon, 1998; Goodwin et al., 2017). Lineage tracing at E9.5 showed that the *Hoxb1* transgene was expressed within the entire *Wnt1-Cre* expression domain (Supplementary Figure 1; Chai et al., 2000; Echelard et al., 1994).

At birth, 8 out of 46 pups died in the first day and had no belly milk spot. These were invariably of the *Wnt1-Cre;CAG-Hoxb1* (mutant) genotype and had a cleft palate (Figure 1a-d). Three had open eyelids. In comparison to their littermates of any other genotype, they had frontonasal hypoplasia and a rounded head, micrognathia, deep-set eyes behind small palpebral fissures and a reduced telencephalon. Although the number of mutants was somewhat lower than the 25% expected, their representation at birth was not significantly reduced (17%; p<0.12). Nonetheless, given the congenital malformations of the survivors, we decided to examine fetal stages, to determine if death also occurred prior to birth.

**Prenatal phenotypes anticipate congenital malformations at two levels of severity**

Between E11.5 and E19.5, Mendelian ratios were respected, and a full quarter of fetuses had a mutant genotype as expected. The signs observed in mutant perinatal pups were present in fetuses older than E15.5, including open eyelids that revealed coloboma of varying degrees of severity, a foreshortened snout, mild frontal bossing with a point at the caudal edge of the telencephalon, mandibular and external ear hypoplasia, and cleft palate (Figure 1e-h).

However, a subset of mutant fetuses observed from E15.5 on demonstrated a far more dramatic phenotype that had not been observed at birth (Figure 2a-h). This included dorsal or ventral exencephaly, frontonasal and maxillary agenesis, and striking mandibular aplasia or agnathia. Remnants of arrayed whisker fields were often visible in the residual ectoderm of what would have become the lateral muzzle, and a small pointed tongue remained present. Severe coloboma or secondary anophthalmia, recognizable from deep remnant retinal pigmented epithelium, characterized all of these mutants (Figure 2b, 2f, 4g-h). No further gross anatomical defects were observed in the skeleton, limbs, adrenal glands or the gut.

Morphological differences distinguished hypomorphic from severe phenotypes at E11.5 as well; the former had both telencephalon and developing nostrils but were visibly smaller than non-mutant littermates (Figure 1i-l), and the latter lacked identifiable growth of the telencephalic vesicles or the frontonasal bud (Figure 2i-l). At E11.5, the maxillary and mandibular buds of the first pharyngeal arch were significantly less developed as measured by width in both mild ($p<0.02$) and severe phenotypes ($p<0.001$) respectively, with a greater effect on the mandible ($p<0.005$ in mild, $p<0.00001$ in severe). Distances measured on photographs between the nostrils, or of the width of the telencephalon, were not significantly different between mildly affected and control littermates. However, vertical growth of the telencephalic vesicles was significantly reduced in even mildly affected mutants ($p<0.01$). Thus, newborn malformations reflected the consequences of a degree of phenotypic variation already present at E11.5, affecting derivatives and tissues adjacent to the frontonasal bud and first pharyngeal arch in all, as well as more posterior pharyngeal arches in some.

## Cardiac defects reflect the differential penetrance of phenotype

Because NC cells of the hindbrain are critical for development of the caudal pharyngeal arches and migrate into the outflow tract of the heart, we examined mutants for cardiovascular malformations. Correlating with the severity of prenatal craniofacial phenotypes, defects were frequently found between E15.5 and birth in terms of misaligned but septated aorta and pulmonary trunk (n=6/13) or as abnormally branched or connected pharyngeal arch arteries (n=5/13; Table 1). At E11.5, five pharyngeal arches have normally developed in controls, with the posterior arches patently connected to the dorsal aorta. However, two hypomorphic mutant embryos developed pericardial edema and patent connections at the level of hypoplasic anterior arches, while others appeared normal (Figure 3a, 3c). Severely affected mutants only developed vascular connections to the level of the second aortic arch (Figure 3b, 3d) and had fully penetrant pericardial edema.

Misalignment of the great arteries was frequent (Figure 3E-F). By birth, severely affected embryos regularly presented pharyngeal arch artery defects such as interrupted aortic arch or abnormal connections between carotid and brachiocephalic arteries (Figure 3G-H). Bicuspid aortic valve being a common, but not always symptomatic, cardiac malformation, we examined valves in histological sections between E15.5 and birth (Table 1). Nearly two-thirds (9/14) of mutants had bicuspid aortic valves (Figure 3i-j), correlated with a dramatic reduction in their NCC component as shown by lineage tracing (Figure 3k-l). Bicuspid valves were observed in both mildly and severely affected mutant animals.

## Differential phenotypes develop between E9.0 and E11.5

### *Mutant NCC follow normal migration patterns in head at E9.5*

Early migratory patterns of NC cells into the pharyngeal arches at E9.5 appeared similar between control *Wnt1-Cre; R26R-LacZ* and mutant *Wnt1-Cre; CAG-Hoxb1; R26R-LacZ* embryos (Figure 4a-b) as well as between *Wnt1-Cre; Rosa*tdT and mutant *Wnt1-Cre; CAG-Hoxb1; Rosa*tdT embryos (Supplementary Figure 1). The dorsal diencephalon and mesenchyme around the dorsal optic vesicles, sometimes appeared less occupied by mutant cells (Figure 4b, asterisk), but this was not distinguishable from individual variation. The frontonasal bud was similarly occupied by mutant or WT NCC at E9.5.

However, by E15.5, *Wnt1-Cre;CAG-Hoxb1;R26R-LacZ* mice consistently showed that NCC had not invested the proximal outflow tract as much as in their control littermates, particularly along the ventral pulmonary trunk (Figure 4c-d). This observation is consistent with the variable cardiovascular defects in mutants and reduced integration of mutant NC cells into the aortic valve, as described above (Table 1 and Figure 3).

**Egr2/Krox20 *and* Fgf8 *expression are not visibly affected at E9.0-E9.5***

*Egr2/Krox20* TF mRNA is normally expressed at E8.5-E9.0 in both rhombomeres r3 and r5, and disappears subsequently from the r3 neuroepithelium, although it remains transcribed in proximal NCC migrating away from both areas at E9.5. As such, *Krox20* is a readout of the organization of rhombomeres r3, r4 and r5 during this critical window of NCC delamination and population of the pharyngeal arches.

The distribution of *Krox20* transcripts seemed unaffected in the rhombomeres and the stream of NCC from r5/r6 towards posterior pharyngeal arches, in mutant (n=11) *versus* control (n=12) embryos from the same litters (Figure 4g-h).

*Fgf8*, encoding an important secreted growth factor needed for multiple processes of craniofacial development, is normally expressed at E9.5 in the developing brain at the midbrain-hindbrain constriction known as the isthmus, as well as at the rostral telencephalic midline (Figure 4i; Trumpp, Depew, Rubenstein, Bishop, & Martin, 1999). These expression domains, as well as an ectodermal domain in the first pharyngeal arch (red arrow), are unperturbed in stage-matched *Wnt1-Cre; CAG-Hoxb1* mutant embryos (Figure 4j; n=5). However, by E13.5 (n=3), normal *Fgf8* expression at the edges of the nasal pits (Figure 4k) was absent from mutants, as was the frontonasal bud between them. Mutant embryos maintained vestigial ectodermal *Fgf8* expression at the edges of the maxillary and mandibular buds, otherwise highly hypoplasic (arrowheads, Figure 4k).

**Severe ocular malformations induced by *Hoxb1*-expressing periocular mesenchyme**

NCC are critical for eye morphogenesis. As mesenchyme, they surround the optic vesicle as it grows towards the lateral ectoderm, contributing such periocular and anterior chamber structures as the matrix of the cornea, the melanocytes of the iris and ciliary body, the pericytes of the choroid, hyaloid and retinal blood vessels, and the sclera. In mice, the choroid is also heavily invested by NCC-derived pericytes, in juxtaposition to the retinal pigmented epithelium (RPE; Chassaing et al., 2016). Mildly affected *Wnt1-Cre;CAG-Hoxb1* fetuses consistently have ocular malformations such as coloboma (Figure 1f), while severely affected animals undergo secondary microphthalmia, showing only rudiments of remaining RPE (Figure 2b). Examination of a *Wnt1-Cre;CAG-Hoxb1;R26R-LacZ* fetus at E18.5 in section shows that NC-accompanied hyaloid vasculature to the lens remains persistent well after its usual period of remodeling into the vascular plexus of the retina. Only small cartilaginous nodules appear to persist of the NC-derived sclera, while corneal NC cells have not stratified and thickened into the appropriate transparent structure (for comparison, adult *Wnt1-Cre;R26R-LacZ* cornea in Figure 4g). Ectopic differentiation of RPE all around the optic vesicle also occurs within the neural retina itself, which subsequently is not layered (arrow, Figure 4h).

## Hypomorphic and severe mutant phenotypes reflect distinct molecular profiles

Both the timing of symptomatic onset and maintenance of rhombomere identity after induction of ectopic *Hoxb1* expression by *Wnt1-Cre* are consistent with an initial effect on the NCC lineage rather than on the dorsal neuroepithelium as such. To orient future work on potential target effector molecules, we examined the relative global expression in the head and heart at E11.5 (*cf.* fragments featured in Figures 1-2, K-L) of transcripts representing candidate molecular pathways that we hypothesized might be affected in these areas by or within NCC mesenchyme. Normalizing to expression levels of the homogeneously and ubiquitously transcribed *Tbp* and *Dhfr*, we examined transcription of genes encoding bone morphogenetic proteins 2 and 4 (*Bmp2* and *Bmp4*); Cyclin D2 (*Ccnd2*) and the cyclin-dependent kinase inhibitor p21 (*Cdkn1a*); the LIM and senescent cell antigen-like domains 1 adhesion protein, also known as Pinch1 (*Lims1*), the homeobox transcription factors *Hoxa2*, *Msx1*, *Pax3*, *Pitx2* and *Sox9*; *Shh* and its principal dependence receptor and transcriptional target *Ptch1*; and the growth factors *Fgf8* and *Vegfa*.

Embryos from two large litters were genotyped and classified into three phenotypic groups: control, hypomorphic (mild) mutant or severe mutant. The relative transcription levels of *Bmp2*, *Bmp4*, *Vegfa*, *Ptch1*, *Lims1*, *Cdkn1a*, *Pax3* and *Sox9* were significantly repressed in the severely affected mutants at E11.5, but not in their mildly affected mutant littermates, as compared with controls. Three of these were in fact significantly *more* transcribed in the mildly affected mutants: *Pax3*, *Sox9* and *Ptch1*, while *Msx1* was unaffected in severe mutants and only somewhat, if significantly, reduced in hypomorphic mutants.

## Discussion

### Ectopic *Hoxb1* expression in NC induces major molecular as well as morphological changes in the developing head, neck and heart

The lack of a significant change in *Ccnd2* associated with a large decrease in *Cdkn1a* transcription is consistent with increased p53-independent apoptosis rather than reduced proliferation in severely affected mutants (Gartel & Tyner, 2002). This cell death may occur in the anterior NCC population forced to express *Hoxb1*, but also in the ocular, craniofacial, endocrine and brain tissues for which its inductive activity is needed. Indeed, older work has shown that experimental removal of NC cells in chick embryos induces brain and pituitary apoptosis, in part from the loss of the perineural vascular plexus, accompanied by NC-derived pericytes from dorsal to ventral forebrain, that will develop into the meninges of these areas (Etchevers, Couly, Vincent, & Le Douarin, 1999).

As supported by changes in the transcription of multiple growth factors, many of the effects of ectopic *Hoxb1* expression in cephalic NC are thus likely to be secondary. Complementary work has shown that *Vegfa* is also an essential growth factor secreted by *Hox*-negative NC mesenchyme, and we have shown here that this gene is less expressed in severely affected *Wnt1-Cre;CAG-Hoxb1* mutant heads; its homozygous loss from the same population leads to cleft palate, reduced ossification of the premaxillary and frontal bones, and maxillary hypoplasia preceded by reduced vascular growth within PA1 (Wiszniak et al., 2015). Vegfa is also chemotactic for NC cells and expressed by the ectoderm of PA2, inciting *Hoxb1*-expressing NC from r4 to populate this arch (McLennan, Teddy, Kasemeier-Kulesa, Romine, & Kulesa, 2010). Ectopic transcription of *Hoxb1* may confer a measure of "r4-origin" identity on the *Hoxa2*-positive cephalic NC of r2/r3 and repress endogenous Vegfa, forcing cells to respond to exogenous signal from PA2 rather than to assume a less directed phenotype that would allow them to better disperse and colonize more rostral tissues (McLennan et al., 2015).

Another chemotactic factor for cephalic NC is Sonic hedgehog, in particular for mesencephalic NC surrounding the optic vesicles (Tolosa, Fernández-Zapico, Battiato, & Rovasio, 2016). Shh is also a survival factor for Hox-negative NCC (Ahlgren & Bronner-Fraser, 1999). As expected, *Shh* expression is not altered in *Wnt1-Cre; CAG-Hoxb1* mutant embryo heads, but that of its dependence receptor and transcriptional target *Ptch1* is. Human *PTCH1* missense mutations are the second most common genetic cause of clinical anophthalmia (Chassaing et al., 2016) and their pathogenic mechanisms are incompletely understood. Shh produced by the foregut endoderm specifically amplifies the sublineage competent to give rise to mesenchymal derivatives such as vascular smooth muscle and skeletal elements in PA1 and 2, as opposed to neurons (Brito, Teillet, & Le Douarin, 2006; Calloni, Glavieux-Pardanaud, Le Douarin, & Dupin, 2007). The Shh source in the ventral diencephalon and optic stalk may act similarly for maxillary and palatine mesenchyme (Tolosa et al., 2016). Thus, effects on *Ptch1* observed in our *Wnt1-Cre; CAG-Hoxb1* mutants implicate Shh among the paracrine exchange of signals between NC-derived periocular mesenchyme and the ocular primordium, as an effector responsible for coloboma and secondary anophthalmia.

*Fgf8* is a known survival factor for PA1 and PA2 mesenchyme (Trumpp et al., 1999). As expected, its transcription in the oral ectoderm was not significantly affected by the ectopic expression of *Hoxb1* in the underlying NC cells of PA1, as assessed both by qRT-PCR and *in situ* hybridization. However, although it did not reach significance, diminution in overall *Fgf8* transcription in the most severely affected embryos at E11.5 likely reflects the secondary loss of the entire frontonasal bud by E13.5 and the nasal pits that also normally express *Fgf8* in their ectoderm, as seen by *in situ* hybridization. Ectopic *Hoxb1* expression leads therefore to both direct and highly indirect transcriptional consequences for facial development, which in the future could be distinguished with an appropriate antibody developed for chromatin immunoprecipitation.

*Hoxb1* overexpression in the NC lineage drastically decreases *Bmp2* and *Bmp4* transcription in severely affected mutants, correlated with a drop in chondrogenic *Sox9* TF transcription. A Shh-soaked bead implanted into the frontonasal process of the chicken embryo not only increases NC-derived mesenchyme but also the relative expression of *Bmp2* and *Bmp4* therein (Hu, Young, Li, Xu, & Hallgr, 2015). Like *Ptch1*, *Sox9* shows a modest but significant increase in transcription in the hypomorphic embryos. These observations seem to implicate defective Shh signal transduction upstream of Bmp expression as effectors of the dramatic facial hypoplasia observed in severe mutant phenotypes.

## Gain of *Hox* function restricts rostral mesenchymal NC competence

Distinct models have induced ectopic expression of paralog group 1 and 2 *Hox* genes in the heads of vertebrates, inducing skeletal but also additional malformations.

The injection of mRNA encoding *Hoxa1* into fertilized eggs of zebrafish prevents the formation of the cartilages of the first pharyngeal arch and thereby, the lower jaw (Alexandre et al., 1996). Such ectopic expression is induced in a mosaic manner in approximately a quarter of all cells, not only NC cells, but also not in all NC cells. Normally, as compared to the conserved expression domain of *Krox20*, the anterior limit of *Hoxa1* expression in the hindbrain is at the r3/r4 boundary, as in other vertebrates. After injection, the hindbrain becomes mispatterned, with enlargement and displacement of r3. Mortality increases to 30% of *Hoxa1*-injected embryos. Survivors of this mosaic *Hoxa1* gain-of-function lose the stream of *Dlx2*-expressing NCC to the 1[st] pharyngeal arch. Subsequently, the jaws do not develop. In addition, the ethmoid cartilage, derived from the frontonasal bud, but also cartilages from the posterior pharyngeal arches are severely reduced or fused, and microphthalmia is also evident. These morphological effects can be phenocopied by exogenous retinoic acid (Alexandre et al., 1996), a known activator of anterior *Hox* gene transcription and teratogenic both in deficiency and in excess.

Spontaneous *Hox* gene overexpression is associated with congenital cranial malformations. A study of late-stage death in over ten thousand commercial broiler chickens *in ovo* showed that calvarial aplasia, leading to exencephaly, accounted for nearly half of the axial skeletal defects and was a frequent birth defect at over 1% of the population. Exencephaly just before hatching correlated with a large increase in absolute and relative *Hoxa1* transcription (Jaszczak, Malewski, Parada, & Malec, 2006). The frontal bones in chicken are NC derivatives, as are their underlying meninges (Couly, Coltey, & Le Douarin, 1993). Supported by our work, regulatory regions of *Hox* paralog group 1 and 2 transcripts make appealing and novel genomic candidates to study in the context of rostral neural tube defects.

Experimentally forcing *Hoxa2* expression in the Hox-negative forebrain and midbrain neural folds of the chick, before NC migration, yielded NC cells that normally colonized the facial primordia at a stage equivalent to mouse E9.5. Later, however, they did not produce upper or lower beak, and the fetuses became exencephalic, with loss of the calvaria and forebrain (Creuzet, Couly, Vincent, & Le Douarin, 2002). NC cells ectopically expressing *Hoxa2* still contributed to the pericytes of facial vasculature, and the eyes were normal. Similar experiments carried out with ectopic *Hoxa3* or *Hoxb4* in *Hox*-negative NC mesenchyme again did not interfere with the initial dispersion of this mesenchyme throughout the frontonasal and PA prominences. Ectopic *Hoxa3* induced severe midfacial hypoplasia, absence of the ocular sclera and loss of the lower jaw but retained a reduced telencephalon. In embryos carrying *Hoxb4*-transfected neural folds, the upper facial skeleton did not form at all, but the proximal portion of the lower jaw did. Co-introduction of exogenous *Hoxa3* and *Hoxa4* combined the most extreme effects of each on the craniofacial skeleton, but the ability of these modified NC to differentiate into Schwann cells of cranial nerves was maintained (Creuzet, Couly, Vincent, & Le Douarin, 2002). The eyes remained exempt from major malformations.

Later work induced *Hoxa2* overexpression in murine NC with a similar approach to that described in this work: *Wnt-Cre* mice were crossed with a floxed *CAG-Hoxa2* allele inserted at the *R26R* locus (Kitazawa et al., 2015). The mutant mice also showed a short snout, exencephaly, cleft palate and apparent coloboma. There were distinct effects of *Hoxa2* gain-of-function on the former *Hox*-expressing NCC derivatives of PA1 compared to the naïve Hox-negative NCC rostral to that. The former group of cells underwent partial mirror image-duplications of hyoid structures in the proximal mandible, while the distal mandible (unlike the rest of the face) was not significantly affected. Similar to our findings, in mutant PA1-2, *Fgf8* transcription was unaffected but *Sox9* was significantly downregulated. PA2 developed malformed but identifiably appropriate hyoid skeletal elements and lost the stapedian artery, which were attributed to the relative cell-autonomous increase in *Hoxa2* compared to other *Hox* genes such as *Hoxb1* (Kitazawa et al., 2015).

The morphological transitions between "new" (rostral) and "old" (caudal) head and body are therefore encoded at a molecular level. The anterior end of the hindbrain beyond the future cerebellum is the source of some *Hox*-negative NC, derived from neuroepithelial levels having once expressed *Hoxa1* and *Hoxa2*, which enter PA1 to form proximal bones of the jaw joint. In contrast, NC cells from the presumptive diencephalon, mesencephalon, isthmus and first rhombomere, never express *Hox* genes before their dispersal throughout the rostral head, and the tail end of this population contributes most mesenchyme to PA1. Caudal *Hoxb1*-expressing NC cells of PA2 form the hyoid bones, while NC mesenchyme of the posterior arches give rise to laryngeal, thyroid and cricoid cartilages as well as shape the branchial vascular network that originates in those arteries splitting off from the ventral aorta and irrigating the face and brain (Etchevers, Vincent, Le Douarin, & Couly, 2001). The structures from NC having once expressed *Hox* genes can undergo "homeotic transformation" (acquisition of a more posterior positional identity in concert with morphological boundaries; Burke, Nelson, Morgan, & Tabin, 1995) and in this, are distinct from the rest of the *Hox*-negative NC that came from regions never having expressed these transcription factors. Here, gain of *Hox* function leads to a homeotic shift in cellular potential relative to the mesectodermal ability to exert a paracrine influence on rostral head structures and to later differentiate into corresponding auxiliary support tissues. The present work introduces evidence that the inductive role of *Hox*-negative cephalic NCC depends also on the continued absence of Hox transcription factors. This role normally precedes vascular remodeling and chondrogenesis, playing a vital role in growth and survival not only of facial bones and teeth but also, secondarily, of eyes, nose, external ears, pituitary, forebrain and facial muscles.

## Loss of *Hox* function affects competence of caudal mesenchymal neural crest

We speculate that the widespread early death in *Hoxa1*-injected gain-of-function zebrafish embryos may be due to an uncharacterized cardiovascular defect in mosaics with greater allele load, since levels of murine *Hoxa1* determines both mesodermal contributions to the outflow tract (Bertrand et al., 2011) and those of the hindbrain NC to the 4[th] pharyngeal (aortic) arch in particular (Roux et al., 2017). *Hoxa1*-null mice, and human patients with truncating mutations of *HOXA1*, also demonstrate cardiovascular malformations similar to, though more penetrant than, those we observed in our *Hoxb1* gain-of-function model (Makki & Capecchi, 2012). They also can be deaf, due to cochlear aplasia, and may have bilateral Duane syndrome, an oculomotor disorder, or central nervous anomalies affecting hindbrain nuclei (facial weakness, central hypoventilation) and intellectual abilities – with a wide phenotypic spread (Tischfield et al., 2005).

*Hoxb1*-null mice have not been reported to have to defects in neural crest-derived skull bones or blood vessels, only in the facial motor nucleus and corresponding portion of facial nerves. Here, too, *Krox20* expression and r4 neural crest cell migration were unaffected in mutants (Goddard et al., 1996). However, some homozygous mutant individuals had died at birth without a belly milkspot, consistent with a potential cleft palate. Skull defects may only have been subtle at the level of the jaw hinge, as in *Hoxa2*-mutant mice (Rijli et al., 1993), but were not examined. Human mutations in *HOXB1* identified to date are missense mutations that, because of their similarities with the null mutant mouse phenotype, have been presumed to be loss-of-function. Here, too, the phenotypic spectrum includes facial nerve palsy and sensorineural hearing loss, but also includes esotropia (which can result from Duane syndrome), low-set ears or external ear malformations, midface retrusion (resulting from underdeveloped maxillary prominences), flat nasal bridge and upturned nose tip, and micrognathia (Vogel et al., 2016; Webb et al., 2012). Like *HOXA1* mutations, *HOXB1* mutations identified to date preferentially affect derivatives of the posterior pharyngeal arches or hindbrain, but have an additional, intriguing effect on the depth of PA1 and midline frontonasal derivatives, without affecting facial width.

## *Hoxb1* genetically interacts with other paralog group 1 and 2 genes to regulate morphogenesis in heart, great vessels and eyes

Recent studies on single and compound *Hoxa1;Hoxb1* mutant embryos showed that reduction of *Hoxa1* gene dosage in a *Hoxb1*-null genetic background results in abnormal PA and heart development and subsequently in great artery defects (Makki & Capecchi, 2012; Roux et al., 2017; Roux, Laforest, Capecchi, Bertrand, & Zaffran, 2015). Lineage tracing analysis shows that both *Hoxa1* and *Hoxb1* genes are expressed in hindbrain NCCs that contribute to PAs 3-6 and outflow tract development (Bertrand et al., 2011; Roux et al., 2017). Endothelial cells in PA4 differentiate normally in compound mutants, but the corresponding vascular smooth muscle is absent. Here, we have shown that NC-targeted *Hoxb1* overexpression also results in defects such as interrupted aortic arch and misalignment of the great arteries. Lineage tracing analysis in this genetic context clearly shows reduced NC colonization of the outflow tract region (see Figure 3b,d), suggesting that cardiovascular defects observed in *Wnt1Cre;CAG-Hoxb1* transgenic embryos is caused by insufficient numbers of NC cells. In the chick, ablation of the cardiac NC population leads to a similar if more severe phenotype along the same spectrum of anomalies (Kirby, Gale, & Stewart, 1983; Waldo, Miyagawa-Tomita,

Kumiski, & Kirby, 1998). However, we never observed an effect on thymus development, as is often associated with cardiac anomalies in ablation experiments.

We also detect NC deficiency at the level of the valve leaflets in *Wnt1-Cre;CAG-Hoxb1* mice, correlated with a bicuspid aortic valve in nearly two-thirds of transgenic fetuses. Arterial valve development is susceptible to imbalances in the ratio of mesodermal to NC cells in the leaflets (Odelin et al., 2014; Phillips et al., 2013). The *Egr2/Krox20*-expressing NC subpopulation emigrating from r5-r6 (see Figure 4g,h) is particularly important for arterial valve development (Odelin et al., 2014, 2018). Thus, ectopic expression of *Hoxb1* in NCCs from r5-r6 may compromise their fate by interfering with the transcriptional readout of other transcription factors such as Krox20 (Desmazières, Charnay, & Gillardi-Hebenstreit, 2009).

The periocular mesenchyme of NC origin is particularly critical for eye development. It gives rise not only to most components of the anterior chamber (cf. Figure 4e): iris, corneal stroma and endothelium, but also to the cartilaginous sclera, the melanocyte-embedded choroid behind the retinal pigmented epithelium, and the smooth muscle and pericytes of the hyaloid and ophthalmic arteries and retinal vascular plexus (reviewed in Creuzet, Vincent, & Couly, 2005). Lineage tracing in mice confirms these contributions (Gage, Rhoades, Prucka, & Hjalt, 2005). NC-targeted ablation of *Pitx2* induces anterior segment defects as well as secondary anophthalmia, similar to the effect of ectopic expression of *Hoxb1* in the same cells (Evans & Gage, 2005). However, *Pitx2* expression levels were not changed in our *Hoxb1* gain-of-function mutants as compared to control embryos, suggesting that these two transcription factors act independently within NC periocular mesenchyme to control its paracrine effects on eye development. Indeed, one such effect of NC-specific *Pitx2* deficiency is to expand the *Pax2*-expressing domain of the future optic nerve within the neuroepithelium at the expense of retinal cell fate, and subsequently at the expense of the RPE (Evans & Gage, 2005). Although NC-targeted *Hoxb1* overexpression also results in secondary anophthalmia, we observed a RPE around the entire ocular globe remnant, incorporation of the lens but obliteration of the pupil by neural retina itself differentiating into RPE, accompanied by a similar incapacity of the periocular mesenchyme to differentiate into the sclera or cornea.

We also consistently observed low-set external ears and loss of the nasal capsule, philtrum and the olfactory bulbs and cortex. These phenotypes emphasize that repression of *Hoxb1*, in addition to that of other paralog group 1 and 2 *Hox* genes, is strictly necessary in anterior cephalic NC for optimal development and subsequent protection of ancestral cephalic sensory functions such as audition, olfaction and vision in vertebrates. This repression and widespread paracrine dependencies necessary for cephalic development may explain the particular sensitivity of the *Hox*-negative NC cell population to either retinoic acid toxicity (Matt, Ghyselink, Wendling, Chambon, & Mark, 2003) or deficiency (e.g. the syndromic microphthalmia due to *STRA6* deletions; Chassaing et al., 2009) *in utero* and their phenotypic overlap. Ultimately, the human embryo must actively repress *Hox* gene activation in rostral cephalic NC cells or risk developing a spectrum of multiple congenital anomalies.

# Methods

## Embryo collection

Timed pregnant mice were euthanized with CO2 at chosen days following embryonic day (E)0.5, determined based on the presence of a vaginal plug after mating. Progeny were analysed at E9.5, E11.5, and individually euthanized by decapitation at E15.5, E16.5, E18.5 and P0 (day of birth). Mice and embryos were genotyped using primers in Table S1.

## Research with vertebrate animals

*Wnt1-Cre* [*H2afv*$^{Tg(Wnt1-cre)11Rth}$], *Rosa*$^{mT/mG}$ [Gt(*ROSA*)26Sor$^{tm4(ACTB-tdTomato,-EGFP)Luo}$], *Rosa*$^{tdT}$ [Gt(*ROSA*)26Sor$^{tm9(CAG-tdTomato)Hze}$] and *Rosa*$^{LacZ}$ [Gt(*ROSA*)26Sor$^{tm1Sor}$] outbred to a Crl:CD-1 background, were purchased from The Jackson Laboratory and Charles River Laboratories respectively, and maintained under authorizations 32-08102012 and 37-08102012 provided by the regional animal care and use committee (C2eA-14) overseeing our institution.

## Generation of the floxed CAG-Hoxb1-IRES2-eGFP transgenic mouse line

Complete cDNA from mouse *Hoxb1* was provided by Dr. Anthony Gavalas (Biomedical Research Foundation of the Academy of Athens, Greece). The coding sequence was cloned into the BamHI site of a pCIG-IRES-eGFP plasmid (Megason & McMahon, 2002). Briefly, pCIG was made by linker insertion of three nuclear localization sequences into pIRES2-EGFP (Clontech) and transferring this insert to pCAGGS (Niwa, Yamamura, & Miyazaki, 1991). The *Hoxb1*-IRES2-*eGFP* fragment was digested with EcoRV-NotI, blunted and cloned into the EcoRV site of a plasmid containing CAG promoter-loxP-CAT-loxP-cloning site-poly(A) (pBSKISecI CAG-CAT-poly(A), a kind gift from Dr. Yumiko Saga, National Institute of Genetics, Japan). The final construct was linearized with SecI and microinjected into fertilized (C57Bl/6xDBA/2) F2 eggs at a concentration of approx. 1ng/µl, using standard techniques. Injected eggs were re-implanted the day after the injection into pseudo-pregnant (C57Bl/6) foster mothers. We obtained offspring from each of the five positive founder lines and subsequently selected the line with the highest expression.

## Lineage tracing and immunofluorescence

After dissection, hearts were fixed in 4% buffered paraformaldehyde during 10 minutes at room temperature, embedded in OCT and cryosectioned at 12 μm. Standard immunofluorescence procedures were used (Ryckebüsch et al., 2010) with anti-CD31 (PECAM) (1:100, BD Pharmingen 01951A), an Alexa Fluor™-555 conjugated goat anti-rat antibody (1:500, Life Technologies), and DAPI (4',6-diamidino-2-phenylindole) was used at 300 nM as a counterstain to distinguish cell nuclei.

Recombined cells were observed using complementary techniques. eGFP is produced by Cre-expressing cells on a *CAG-Hoxb1* or *Rosa$^{mT/mG}$* background and was visualized, like for immunofluorescence, in cryostat section at 12 μm on a Zeiss Apotome fluorescent microscope. In order to use light microscopy in whole mount, we used the floxed *R26R-LacZ* reporter allele as an additional lineage indicator of Cre recombination. We compared embryos at the same numbers of somite pairs +/- 1 pair. β-galactosidase is produced by Cre-expressing cells and precipitates an insoluble dark turquoise-blue chromogen in the presence of X-Gal (5-bromo-4-chloro-3-indolyl-D-galactopyranoside), as described (Gierut, Jacks, & Haigis, 2014). Whole-mount photographs were taken on a Zeiss Axiozoom dissecting microscope. The head and heart of a *Wnt1-Cre;CAG-Hoxb1;R26R-LacZ* embryo at E18.5 were embedded, sectioned at 10 μm on a cryostat, developed in X-Gal, and counterstained with hematoxylin. To study aortic valves between E15.5 and birth, hearts were sectioned at 10 μm and alternate sections stained with hematoxylin-eosin according to standard protocols. Microphotographs were taken with a Zeiss AxioScan.Z1 slide scanner.

## *In situ* hybridization

Whole-mount *in situ* hybridization was carried out according to standard protocols for mouse embryos at E9.5 and E13.5 (Trumpp et al., 1999). The *Krox20* probe template plasmid was linearized with BamHI and transcribed with T3 RNA polymerase; the *Fgf8* probe template plasmid was linearized with BglII and transcribed with SP6 RNA polymerase.

## qRT-PCR and statistics

Groups of 3 WT (including carriers of individual *Wnt1-Cre* or floxed *CAG-Hoxb1* alleles) or 4 mildly or severely affected mutant embryos were constituted after individual qRT-PCR analysis of the expression levels of 16 genes (primers in Table S1 of the online supplementary materials). Total RNA was extracted using the NucleoSpin RNA extraction kit (Macherey-Nagel). First-strand cDNA was reverse transcribed from 500 ng total RNA with the AffinityScript kit (Agilent) using a mix of random hexamer and oligo-dT primers, and standard qPCR with the LightCycler 480 SYBR Green I Master mix (Roche) carried out on the corresponding LightCycler, according to manufacturers' recommendations.

Expression was normalized to the geometric mean of *Tbp* and *Dhfr* expression in each group. One-way analysis of variance demonstrated that these groups were differentially variable, such that means of the mildly or severely affected groups were compared individually to the control group, using the two-tailed Student's t-test under conditions of unequal sample size and variance, and a significance threshold of $p < 0.03$. Measurements of distances on frontal/ventral photographs of these embryos before lysis (examples shown in Figures 1 and 2, k-l) were taken using ImageJ 1.51w and groups of three of each genotype were compared using a two-tailed Student's t-test under conditions of equal sample size and variance, with a significance threshold of $p < 0.03$.

# Acknowledgments

We thank Elise Plaindoux, Nathalie Eudes and Adeline Spiga-Ghata for their precious technical assistance and Brigitte Laforest and Emilie Faure for helpful discussions. The following colleagues kindly provided probe template plasmids: Piotr Topilko (*Krox20*) and Robert Kelly (*Fgf8*). This work was supported by an Association Française contre les Myopathies Strategic Project award to SZ and postdoctoral fellowships from the Fondation Lefoulon-Delalande to GO, SS and FL.

# Figure legends

## Figure 1.

Comparison of hypomorphic *Wnt1-Cre; CAG-Hoxb1* mutant phenotypes with control mice at birth (P0) and embryonic days (E)15.5 and 11.5. (**a**) Head of newborn mouse carrying the unfloxed *CAG-Hoxb1* transgenic allele. (**b**) Head of mutant littermate, dead at birth, showing microphthalmia, micrognathia, microtia, a foreshortened snout and rounded forehead. Cleft palate, never observed in controls (**c**), is fully penetrant in mutants (arrow). (**d**) Similar phenotypes are observed at E15.5, with a change in the sigmoid curvature of the profile of the frontal region from controls (**e**) to mutants (**f**) allowing for reproducible prediction of genotype, in addition to coloboma or microphthalmia in the latter. (**g**) Control palate at E15.5. (**h**) Cleft palate in mutants (arrow). (**i**) Body size in control embryos at E11.5 is generally larger than (**j**) mutant embryos, a difference no longer observed at E15.5. The maxillary portion of pharyngeal arch (PA) 1 below the eye as well as the other PAs appears hypoplasic. The telencephalic vesicles are visibly smaller than in controls. (**k**) Frontal view of control facial region, truncated at the level of the caudal hindbrain and atrial end of heart. (**l**) Frontal view of mutant facial region at equivalent level. The telencephalic vesicles are no different in width but have grown significantly less coronally. The medial frontonasal bud, maxillary and mandibular (dotted lines) portions of PA1, and PA2, are measurably hypoplasic, while the heart tube appears unaffected. Scale bars: **a-f**, 2 mm; **g-h**, 1 mm; **i-l**, 0.5 mm.

## Figure 2.

Comparison of severe *Wnt1-Cre; CAG-Hoxb1* mutant phenotypes with control mice at birth (P0) and embryonic days (E)18.5, 15.5 and 11.5. (**a**) Head of mouse at late E18.5, carrying the unfloxed *CAG-Hoxb1* transgenic allele. (**b**) Head of mutant littermate, showing secondary anophthalmia, extreme micrognathia, anotia, absence of midface structures and forebrain, persistence of lateral ectodermal vestiges and a small tongue. Cleft palate, never observed in controls at birth (**c**), is severe to the point of agenesis in mutants (**d**), in continuity with a large hole at the level of the presphenoid (arrow). (**e**)

Control embryo at E15.5. (**f**) Mutant embryo at E15.5. Some of these severely affected embryos show dorsal exencephaly as in (**b**), while others have no oral structures below the prospective hypothalamus, which then protrudes below the vestiges of the eyes. The dorsal brain is always truncated at the level of the midbrain vesicles, and the midface and upper jaw are aplasic. (**g**) A ventral view of a control palate at E15.5. (**h**) Palatal agenesis is frequent in mutants, leaving residual lateral processes and incomplete closure of the sphenoid (arrow). (**i**) Body size in control embryos at E11.5 is larger than (**j**) severe mutant phenotypes, a difference no longer observed at older stages and consistent with some minimal selection against the most severely affected embryos. These have pericardial edema and small body somites. The maxillary portion of pharyngeal arch (PA) 1 below the eye is absent and the mandibular portion is severely hypoplasic (dotted lines in k, l) with no caudal PAs visible beyond occasional PA2. The telencephalic vesicles are always absent. (**k**) Frontal/ventral view of control facial region, truncated at the level of the caudal hindbrain and atrial end of heart. Blood is visible in PA3-6. (**l**) Frontal/ventral view of mutant facial region at equivalent level. The ventral diencephalon emerges below the optic vesicles, as no frontonasal NC mesenchyme or telencephalon are present. Lateral vestiges of the maxillary buds are present as well as a hypoplasic mandibular process, with the cardiac outflow tract feeding directly into arteries of PA1 or sometimes PA2. The medial frontonasal bud, maxillary and mandibular portions of PA1, and PA2, are hypomorphic, while the heart tube appears unaffected. Scale bars: **a-c, e-f**, 2 mm; **d**, **g-h**, 1 mm; **i-l**, 0.5 mm.

## Figure 3.

Cardiovascular phenotypes in *Wnt1-Cre; CAG-Hoxb1* mutant mice. (**a**) Mutant embryo at E11.5 with hypomorphic phenotype, frontal/ventral view, truncated after heart. The heart is normally shaped and loops from atria, behind, to left to right ventricular domains (LV and RV respectively) then to the outflow tract (oft) and pharyngeal arches (PA)4-6. (**b**) Entire mutant embryo at E11.5 with severe phenotype, frontal/ventral view, same scale. The heart tube is unlooped and in the absence of an identifiable outflow tract, connected to the sole residual first pharyngeal artery. Left lateral views of (**c**) control and (**d**) severe mutant embryos at E11.5, different individuals from (**a**,**b**), showing phenotypic variability. (**e**) Heart of control embryo at E15.5, showing normal morphology of right and left atria (RA, LA), right and left ventricles (RV, LV) aorta (Ao) and pulmonary trunk (Pt). (**f**) Mutant embryos frequently show misalignment of the great arteries at E15.5 (arrows at level of double outlet right ventricle). (**g**) At birth, the aorta normally emerges from the control heart and arches around the pulmonary trunk, with proximal branches for the brachiocephalic (right), left common carotid and subclavian arteries. (**h**) This mutant, dead at birth, had an interrupted aortic arch – others show more distal aberrant connectivity at the level of the brachiocephalic or common carotid arteries, and some are phenotypically normal. (**i**) Hematoxylin-eosin stain of a transverse section at the level of the normally tricuspid aortic valve in a control embryo at E18.5, showing right coronary (RC), left coronary (LC) and non-coronary (NC) leaflets. (**j**) Similarly processed section at the level of a bicuspid aortic valve in a *Wnt1-Cre; CAG-Hoxb1* mutant at E18.5 (cf. **Table 1**). The left and non-coronary leaflets have fused (NC-LC), evoking the rare human type 3 BAV. (**k**) Cre-expressing neural crest cells (GFP, green using the *Rosa^{mT/mG}* reporter) in the three aortic leaflets of a control embryo at E18.5. (**l**) Cre-expressing neural crest cells (GFP, green) in the three aortic leaflets of a *Wnt1-Cre; CAG-Hoxb1* mutant embryo at E18.5. Immunofluorescence against platelet endothelial cell adhesion molecule (PECAM/CD31) in red outlines the endocardial interface of the leaflets and vascular endothelium in **k-l**. Scale bars: **a-h**, 500 μm; **i-l**, 100 μm.

## Figure 4.

Tracking NC cells in the head, neck and hearts of *Wnt1-Cre* (control) or *Wnt1-Cre; CAG-Hoxb1* (mutant) mice. (**a**) Lineage tracing with the *R26R-LacZ* allele shows ß-galactosidase activity in the dorsal neural tube, and abundant labeled NC mesenchyme in the frontonasal process, around the midbrain, and in the pharyngeal arch (PA) primordia, among other sites. (**b**) No morphological differences can be distinguished between control and mutant embryos at E9.5. Fewer labeled cells appear to be in the dorsal diencephalic midline or to occupy the dorsal periocular mesenchyme (asterisk), but abundant cells are present in the frontonasal bud. (**c**) At E15.5, labeled NC cells in control mice are visible in the smooth muscle walls to the myocardial base of the ventral aorta and pulmonary trunk, the aortic arch and cephalic arteries branching therefrom. (**d**) At E15.5, labeled NC cells in mutant mice are not as proximal to the heart along the aorta as controls, and are even more conspicuously absent from the pulmonary trunk beyond the ductus arteriosus. The great arteries are also misaligned and incompletely rotated. (**e**) Whole-mount view through the eye of an adult *Wnt1-Cre; Rosa*^tdT mouse, showing in red the thick NC-derived stroma of the cornea over the iris, forming the anterior chamber, and smooth muscle along the hyaloid artery (ha) behind the lens. (**f**) Section through the eye of a *Wnt1-Cre; CAG-Hoxb1; R26R-LacZ* mouse, showing differentiation of retinal tissue all around an invaginated and hypoplasic lens, which continues to be supported posteriorly by NC-accompanied hyaloid vasculature (turquoise). The disorganized neural retina is enclosed in retinal pigmented epithelium (RPE), but also shows differentiation of RPE *in situ* amidst the neuroepithelium (red arrow). The cornea is reduced to a monolayer on the external face of the ocular primordium. (**g**) Whole-mount *in situ* hybridization (WISH) against *Egr2/Krox20* transcripts, at E9.5 normally transcribed at higher levels in rhombomere (r)5 than in r3 and in NC cells migrating from the r5/r6 border towards PA3, in a control embryo, dorsal view. (**h**) *Krox20* is expressed identically in stage-matched mutant embryos. (**i**) Fgf8 WISH in a control embryo at E9.5, right lateral view. (**j**) Fgf8 in mutant littermates shows the same distribution in the isthmus (black arrowhead), the rostral telencephalic midline (between arrows) and the oral ectoderm (red arrowhead). (**k**) By E13.5, Fgf8 is normally robustly expressed in facial ectoderm of the nasal pits, the lateral maxillary (mx) and mandibular (md) bud ridges, frontal view. (**l**) In mutant littermates, facial Fgf8 expression is found in vestiges of PA1 ectoderm (arrowheads) and remains expressed to a lesser degree in the diencephalic ventral midline.  Abbreviations: ha, hyaloid artery; LA, left atrium; LV, left ventricle; mx, maxillary bud; md, mandibular bud; n, nasal pits; oft, outflow tract; om, oculomotor muscle; ov, optic vesicle; r3/r5, rhombomeres 3/5; RA, right atrium; RV, right ventricle.

Figure 5.

Relative gene expression in heads and hearts of E11.5 control (grey), mildly affected (blue) and severely affected (pink) *Wnt1-Cre;CAG-Hoxb1* mutant embryos, where fold-change is log(10) transformed and directionality maintained on the plot. Standard deviations portrayed indiscriminately with solid or dashed lines to improve visibility. Primers in Table S1. *, $p < 0.03$; **, $p < 0.01$; ***, $p < 0.001$.

Table 1, Supplementary Figure 1 and Table S1 are separate files

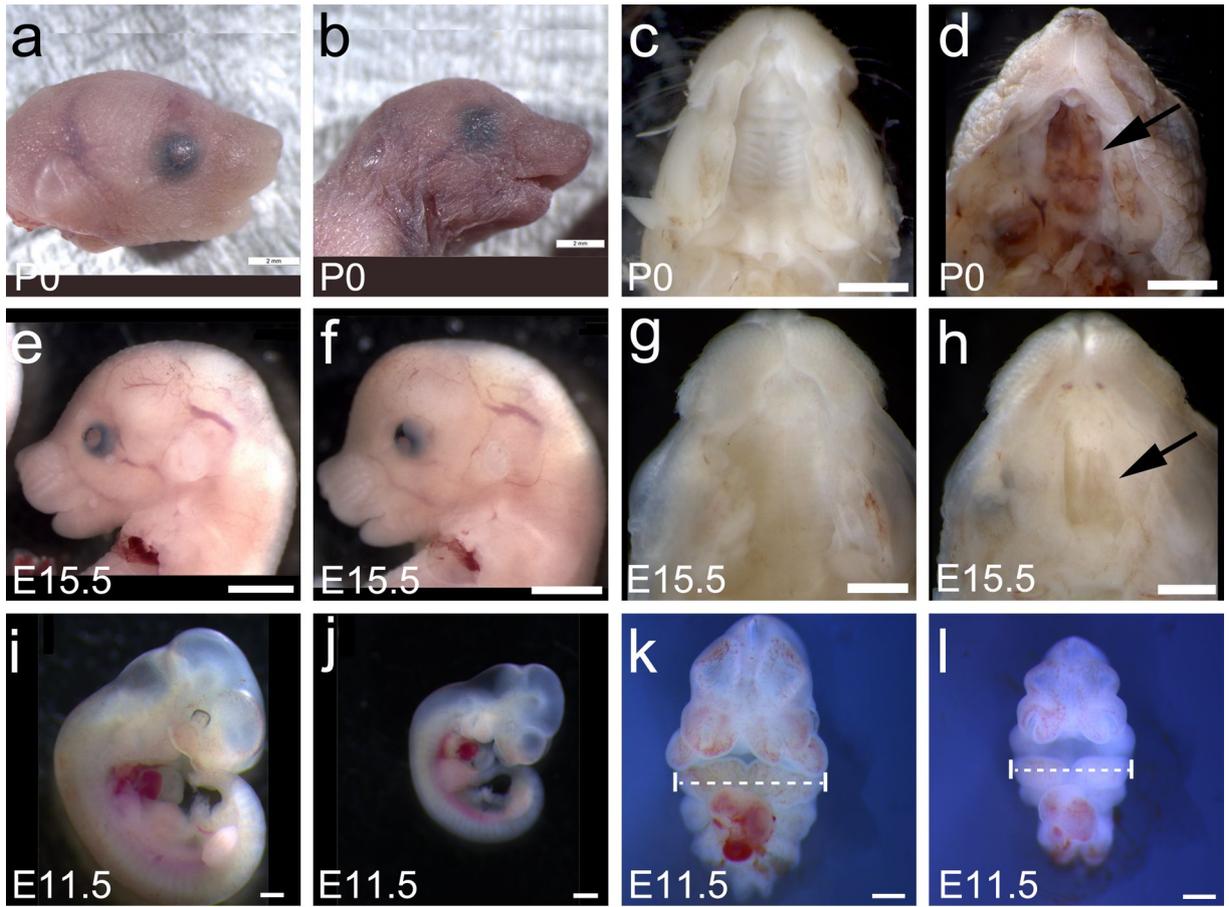

Figure 1

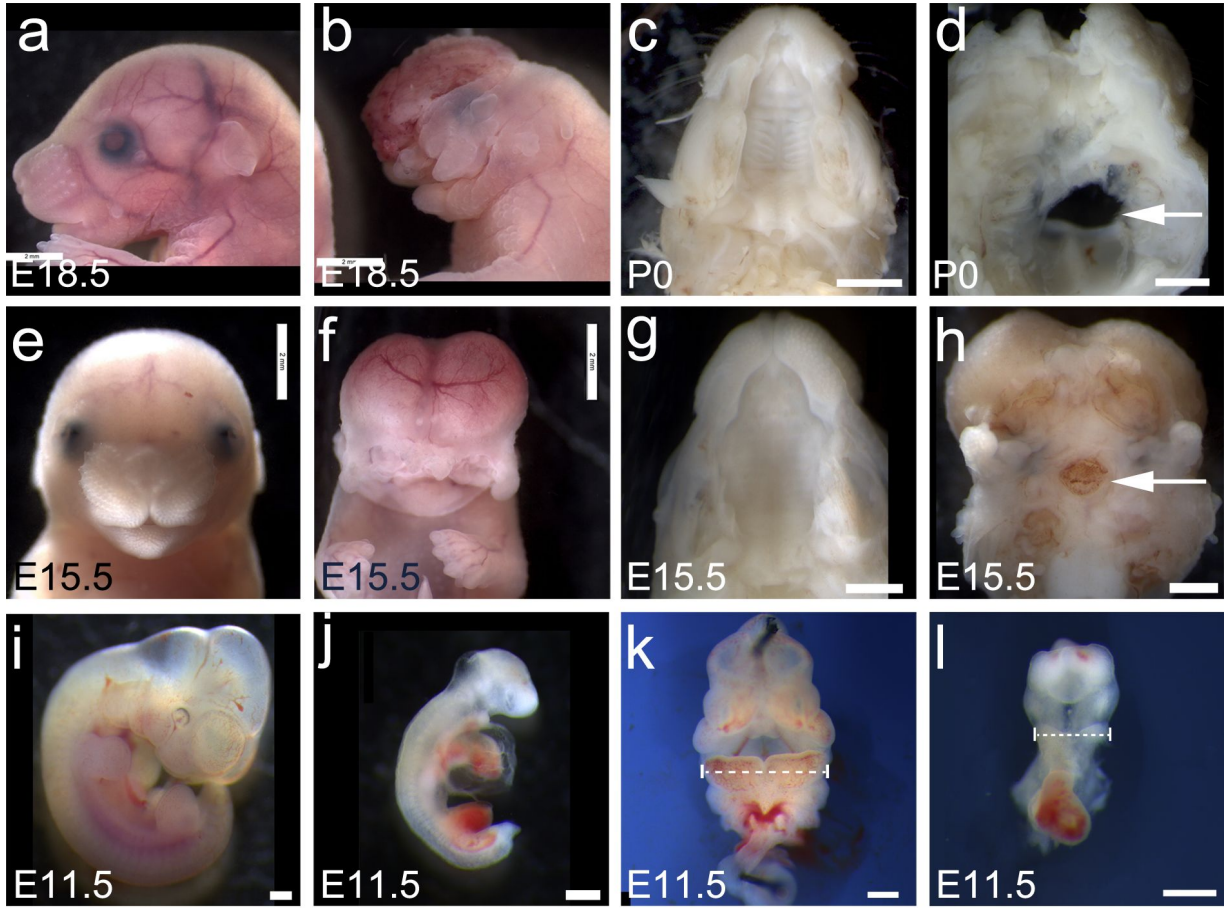

Figure 2

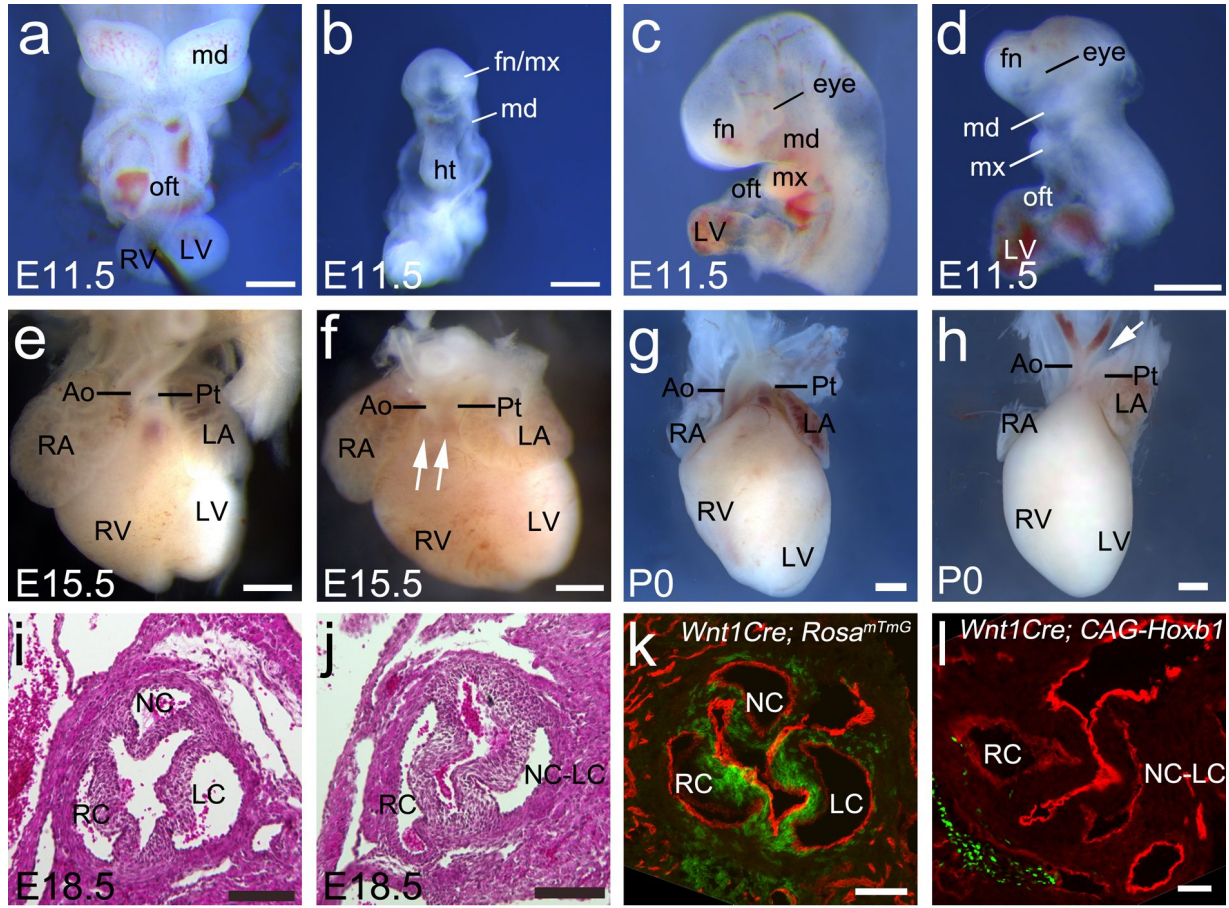

Figure 3

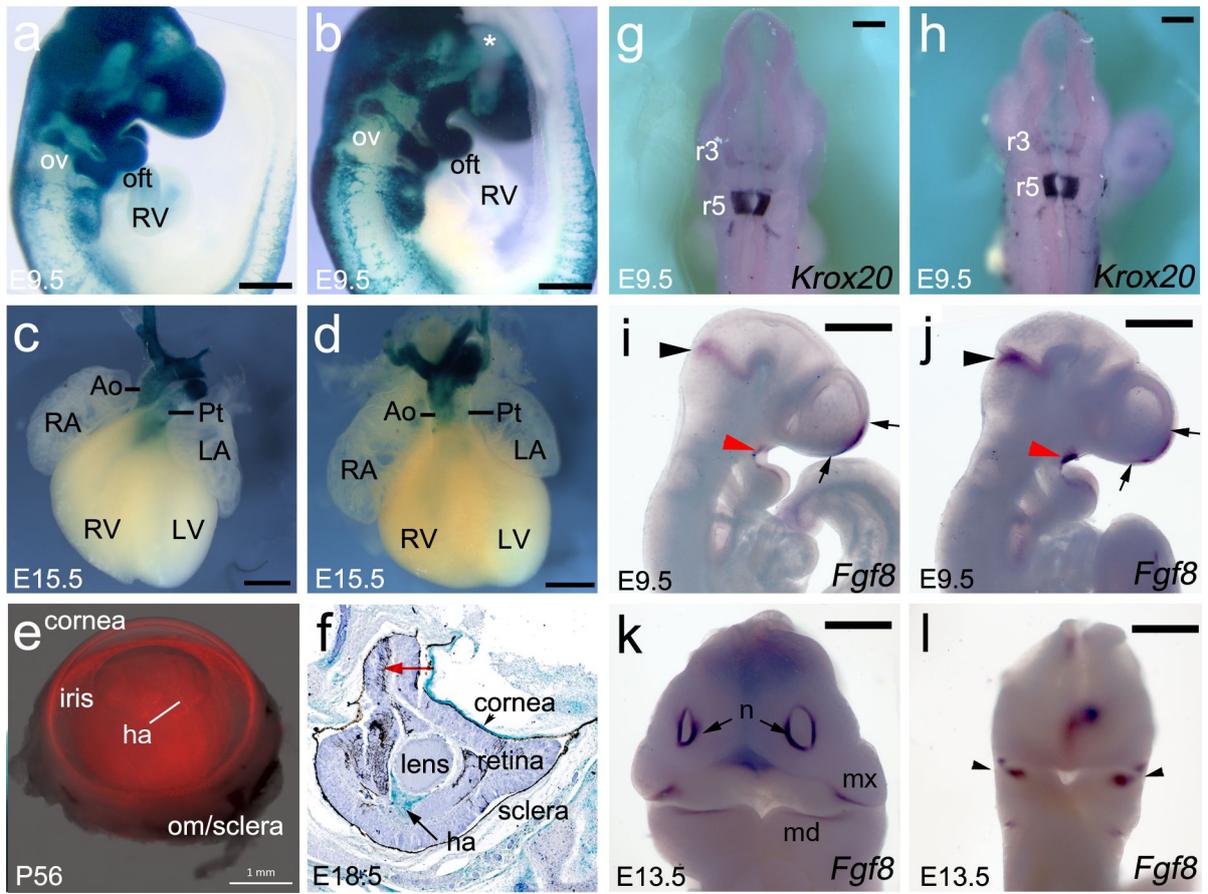

**a** E9.5 ov oft RV

**b** E9.5 ov oft RV *

**g** E9.5 r3 r5 *Krox20*

**h** E9.5 r3 r5 *Krox20*

**c** E15.5 Ao RA Pt LA RV LV

**d** E15.5 Ao RA Pt LA RV LV

**i** E9.5 *Fgf8*

**j** E9.5 *Fgf8*

**e** P56 cornea iris ha om/sclera 1 mm

**f** E18.5 cornea lens retina sclera ha

**k** E13.5 n mx md *Fgf8*

**l** E13.5 *Fgf8*

Figure 4

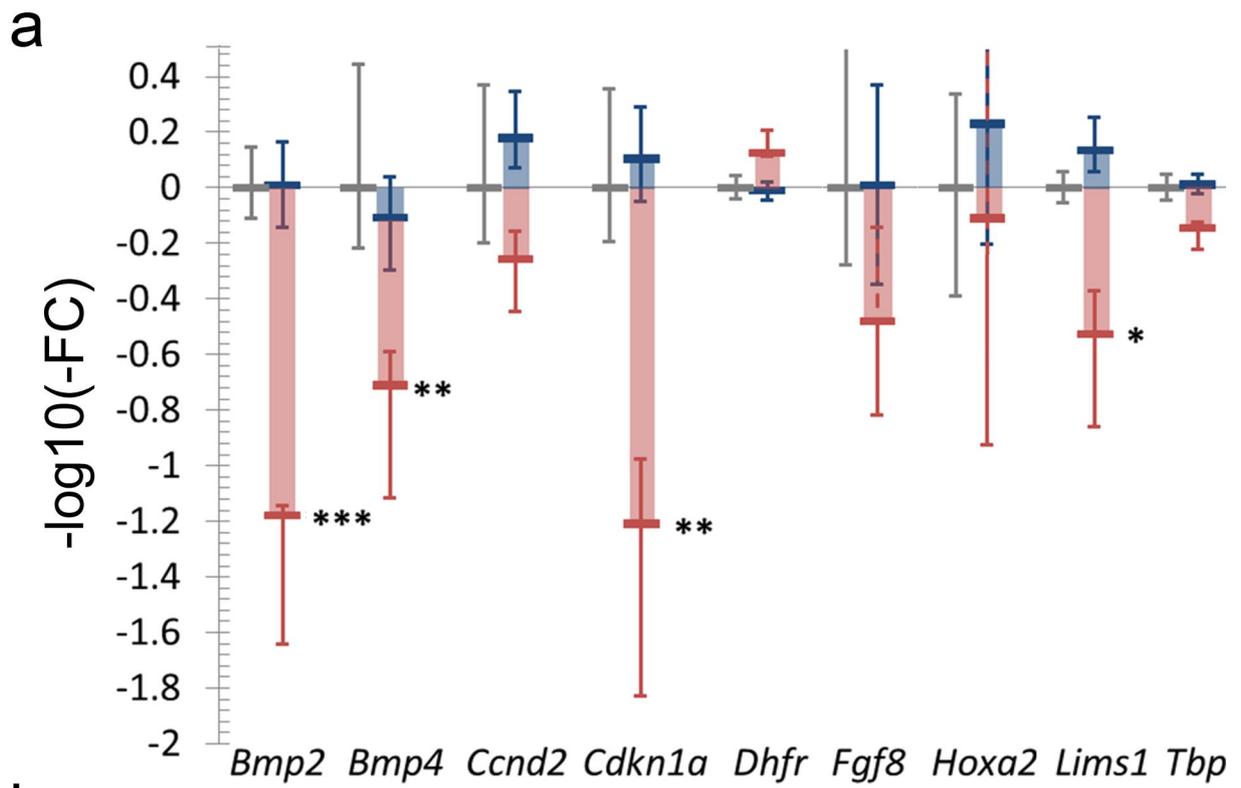

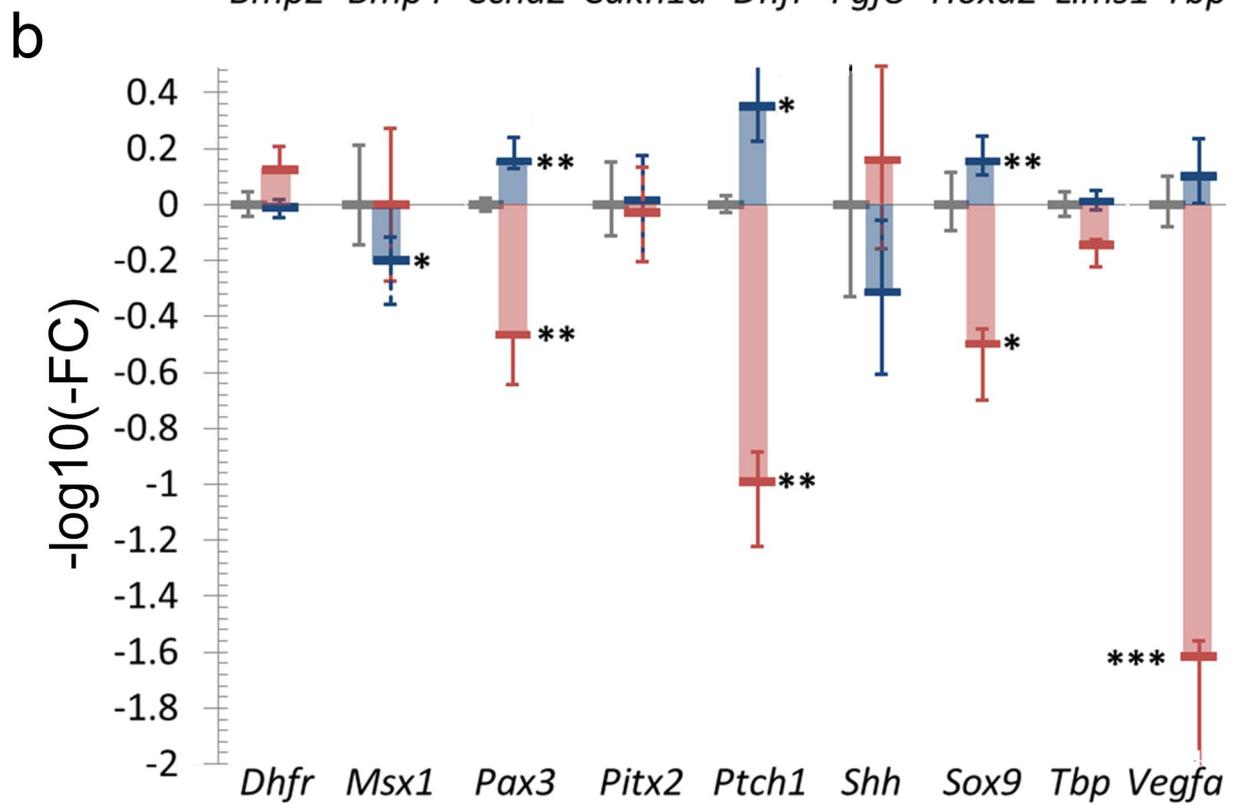

# Figure 5

## Supplementary Figure 1.

**A**. A transgenic construct was designed with the CAG promoter (**C**MV enhancer/chicken beta-**a**ctin promoter/splice acceptor from the rabbit beta-**g**lobin gene), a loxP-flanked STOP sequence (Chloramphenicol Acetyl Transferase [CAT] followed by a SV40 polyadenylation signal), the entire mouse *Hoxb1* cDNA (900bp), an internal ribosomal entry site (IRES), and the cDNA of the enhanced green fluorescent protein (*eGFP*) gene. Prior to Cre-mediated excision of the "floxed" STOP sequence, no ectopic expression of *Hoxb1* is observed in cells. When bred to Cre recombinase-expressing mice, the STOP sequence is removed in Cre-expressing cells of the resulting offspring, allowing transcription of both *Hoxb1* and *eGFP*. **B**. Schematic vertebrate embryo at approximately mouse E9.5 representing neural crest colonization of the face, heart, foregut and somites in red (first panel); the typical *Hoxb1* expression domain in green in rhombomere 4, neural crest cells colonizing the second pharyngeal arch and multiple tissues caudal to the third pharyngeal arch (second panel); and the superposition of transgenic and endogenous *Hoxb1* expression in *Wnt1-Cre; CAG-Hoxb1-GFP* mutant embryos, inducing ectopic expression in the roofplate, isthmus and neural crest cells investing the head, cardiac outflow tract and face. **C**. Reporter (red, Tomato fluorescence) expression at E9.5 after recombination by *Wnt1-Cre*. **D**. Reporter (red, Tomato fluorescence) expression at E9.5 in *Wnt1-Cre; CAG-Hoxb1-GFP* mutant embryo. **D'**. Expression of GFP reporter for the *Hoxb1* transgene in same embryo.

A

Wnt1Cre (+) mouse                          Hoxb1 (+) mouse

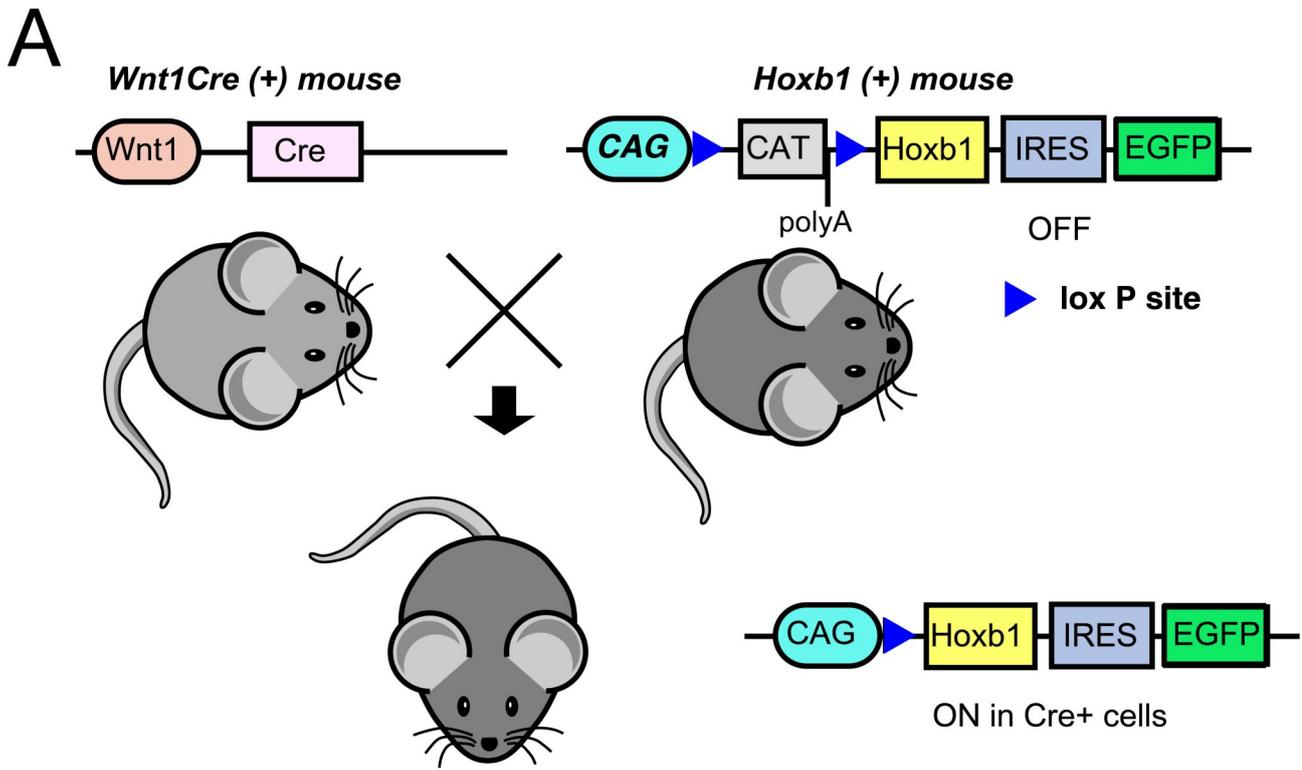

B

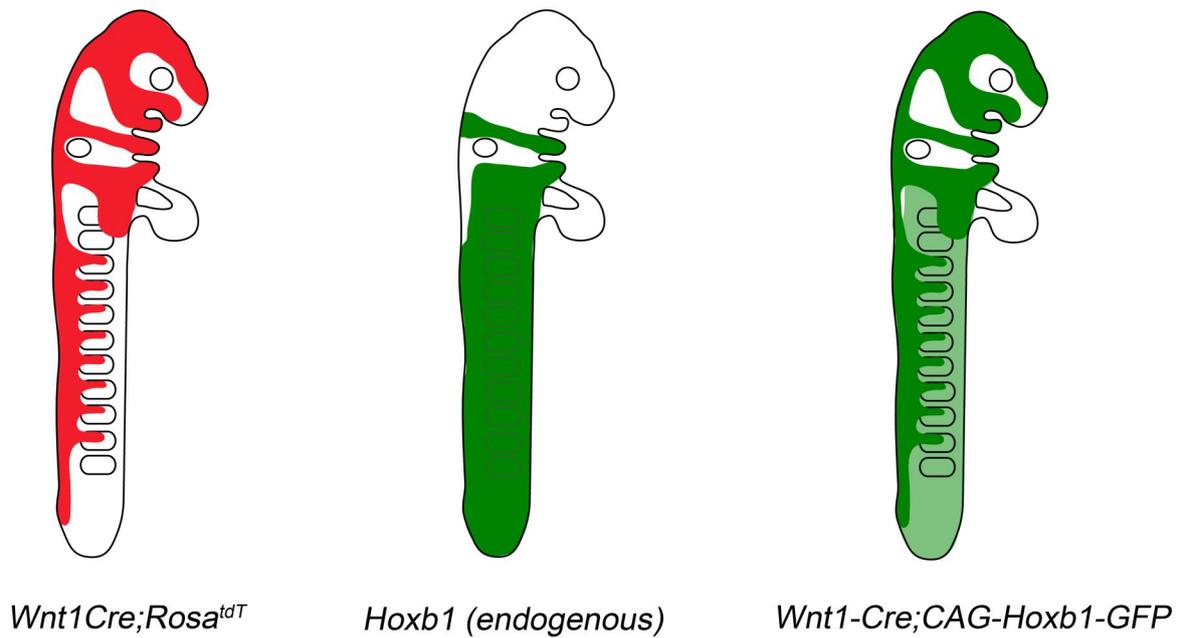

Wnt1Cre;Rosa^tdT          Hoxb1 (endogenous)          Wnt1-Cre;CAG-Hoxb1-GFP

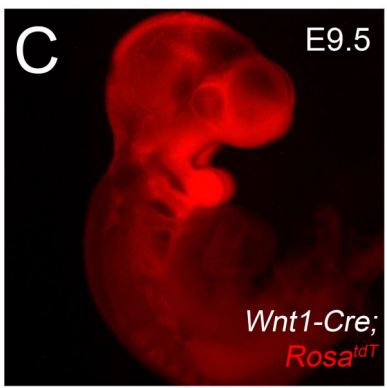

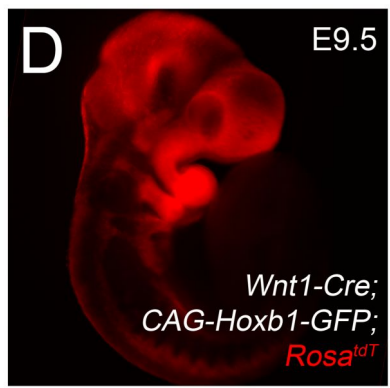

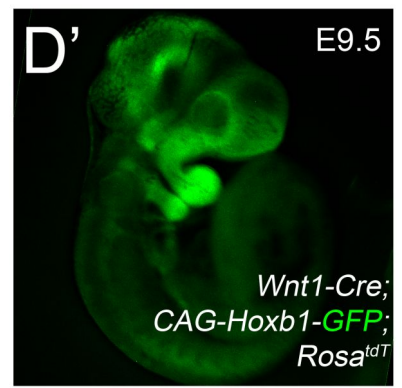

| Stage | CV defects, controls | PAA defects | Arterial trunk defects | PE with PAA defects | No CV defect | Valve defects, controls | BAV | Small NC leaflet | TAV |
|---|---|---|---|---|---|---|---|---|---|
| E11.5 | 0/8 (0%) | **5/15 (33%)** | | **7/15 (47%)** | 3/15 (20%) | n/a | n/a | n/a | n/a |
| E15.5 | 0/5 (0%) | **2/8 (25%)** | **5/8 (62%)** | | 1/8 (13%) | 0/2 (0%) | **2/3 (67%)** | **1/3 (33%)** | 0/3 (0%) |
| E16.5-birth | 0/4 (0%) | **3/5 (60%)** | **1/5 (20%)** | | 1/5 (20%) | 0/2 (0%) | **4/11 (36%)** | **2/11 (18%)** | 5/11 (45%) |

**Table 1**. Cardiovascular (CV) malformations observed in a subset of *Wnt1-Cre;CAG-Hoxb1* or control littermates. PAA, pharyngeal arch artery (comprises interrupted aortic arch); arterial trunk defects include double-outlet right ventricle and over-riding aorta; PE, pericardial edema; BAV, bicuspid aortic valve; NC, non-coronary

## Table S1. Primers for genotyping and qRT-PCR in mice, used in this work

| Transcripts | 5' | 3' |
|---|---|---|
| *Bmp2* | TAAACCGTCTTGGAGCCTGC | TCCGAATGGCACTACGGAAT |
| *Bmp4* | TGCTTCTTAGACGGACTGCG | CTGGGGAAGCAGCAACACTA |
| *Ccnd2* | GACCTTCATCGCTCTGTGC | ATCCTGCTGAAGCCCACA |
| *Cdkn1a* | GCAGACCAGCCTGACAGATT | GAGGGCTAAGGCCGAAGA |
| *Dhfr* | TTTATCCCCGCTGCCATCAT | ACCAGATTCTGTTTACCTTCCACT |
| *Fgf8* | GCTGAGCTGCCTGCTGTT | AGCTCGGAGCAGGGAAGT |
| *Hoxa2* | GAAGGCGGCCAAGAAAAC | CATCAGCTATTTCCAGGGATTC |
| *Lims1* | AAATGCCATGCCATCATTG | CCGAGCATCAGCAGTTAGC |
| *Msx1* | TCAAGCTGCCAGAAGATGCT | GGGACTCAGCCGTCTGGC |
| *Pax3* | ACTACCCAGACATTTACACCAGG | AATGAGATGGTTGAAAGCCATCAG |
| *Pitx2* | CCTTACGGAAGCCCGAGT | AAAGCCATTCTTGCACAGC |
| *Ptch1* | CTTCGCTCTGGAGCAGATTT | TCTTTTGAATGTAACAACCCAGTT |
| *Shh* | TCCACTGTTCTGTGAAAGCAG | GGGACGTAAGTCCTTCACCA |
| *Sox9* | TCGGTGAAGAACGGACAAGC | TGAGATTGCCCAGAGTGCTCG |
| *Tbp* | CCCCACAACTCTTCCATTCT | GCAGGAGTGATAGGGGTCAT |
| *Vegfa* | TTAAACGAACGTACTTGCAGATG | AGAGGTCTGGTTCCCGAAA |

| Transgenes | 5' | 3' |
|---|---|---|
| *Cre* | GGCGCGGCAACACCATTTTT | CCGGGCTGCCACGACCAAG |
| *CAG-Hoxb1* | GCCCAGTTCCATCACCTCTT | GACCAGGATGGGCACCAC |
| *LacZ* | GCATCGAGCTGGGTAATAAGCGTTGGCAAT | GACACCAGACCAACTGGTAGCGAC |
| *GFP* | GAAGCAGCACGACTTCTTCAAG | ATTCTAGACTACAGCTCGTCCATGCCGAGAGC |